# The use of multiple models within an organisation

Chris Dent (School of Mathematics, University of Edinburgh; Turing Fellow, Alan Turing Institute)

Michael Goldstein (Department of Mathematical Sciences, Durham University)

Andrew Wright (Independent consultant; Professor in Practice, Durham University)

Henry Wynn (Centre for Analysis of Time Series, LSE; Alan Turing Institute)

Supported by the Tools, Practices, Systems programme at the Alan Turing Institute.

The authors can be contacted through chris.dent@ed.ac.uk.

# 1. Introduction

Organisations, whether in government, industry or commerce, are required to make decisions in a complex and uncertain environment.

These decisions can be critical to the future of the organisations and its stakeholders. In the case of government, decisions can impact the lives of many millions of people. The way in which models are used to support decision making is a matter of considerable public interest, particularly at a time when the legitimacy of public and private institutions, and "experts" more generally, is being questioned.

The way models are used is intimately connected to the way organisations make decisions and the context in which they make them. Decisions represent choices between alternative courses of action, and the decision maker needs to assess which of these options is likely to result in the best outcome, however that is defined. These decisions need to be made under conditions of uncertainty, and the management of this uncertainty is an important consideration. Computer models, used properly, can help with this assessment.

Typically, in a complex organisation, multiple related models will often be used in support of a decision. For example, engineering models might be combined with financial models and macro-economic models in order to decide whether to invest in new production capability. Different parts of a complex organisation might operate their own related models which might then be presented to a central decision maker. Organisations often use external advisors for decision support and modelling work, whose findings may then be compared or combined with internal modelling exercises. This gives rise to a number of issues (both technical and organisational).

Solutions to these challenges will need to be scientifically and organisationally sound and practically implementable given typically available skills, budgets, timescales and practices. Addressing the challenges of multiple models will require both technical statistical approaches as well as organisational approaches.

Yet in practice, there is little awareness of the practical challenges of using models in a robust way to support decision making. There is significant scope to improve decision making though an enhanced understanding of the role and limitations of modelling and through the application of cutting edge methodologies and organisational best practice.

This report is in the form of a "white paper". Its purpose is to identify key issues for consideration whilst postulating tentative approaches to these issues that might be worthy of further exploration, focussing on both technical and organisational aspects.

For the technical statistical aspects of combining evidence from different computer models, we will consider the use of Bayesian graphical methods. We will also consider the use of emulation for uncertainty quantification for more complex models (including propagation of uncertainty between models).

Organisationally, we will consider whether multiple related, but competing models is a strength or a problem in an organisation, and the best way to manage some of the conflicting evidence that might be provided. We will also consider issues around documentation, curation and repeatability, particularly where proprietary external models are being used. We will consider how complex modelling results can be communicated, particularly in relation to the communication of risk and uncertainty to decision makers that do not have technical expertise in modelling or statistics.

A further key part of this work is at the interface of the technical and organisational. We will consider collaboration between technical modellers and relevant social scientists on how modelling evidence sits within overall decision processes, and how this integration of scientific evidence into decision making can be improved.

The report begins with a framework for consideration of how model-based decisions are made in organisations. It then looks more closely at the questions of uncertainty and multiple models. It then postulates some technical statistical and organisational approaches for managing some of these issues. Finally, it considers the way forward, and the possible focus for further work.

# 2. Model-based decision making in organisations

Organisations, whether public or private sector, social or commercial, will often use models to support their decision-making. The decisions that organisations make can range from short-term operational choices to long-term strategic decisions and everything in between. However, all of these decisions will be in pursuit of the organisation's purpose.

## Organisations and their purposes

An organisation can be defined as a group of people that come together for a specific purpose. The purpose of an organisation is a critical consideration for its decision making. All decisions should be made in support of that purpose. An organisation's purpose will vary by type of organisation.

For a commercial organisation, this purpose is usually to make money for its investors (or more broadly, the creation of shareholder value). Often this is realised indirectly through some intermediate objective, or mission, such a market share or cost optimisation.

Charitable organisations or social enterprises will have their own specific objectives which define their purpose.

The purpose of government is the subject of political and constitutional theory but could be broadly defined as delivering benefits for its citizens. These benefits are typically multi-faceted and will often be subjective. The UK Treasury "Green Book" talks about the creation of "social value" as being the goal of public policy, which is defined as the sum of all costs and benefits to UK citizens, including those that are non monetisable and non-quantifiable.

The same is broadly true for non-governmental public organisations, such as health authorities, universities and regulators, although the scope of responsibilities and powers of these organisations, which almost always derive from government, will be limited.

All organisations operate within the legal, regulatory, political and ethical framework of wider society, which will constrain and influence their choices. Organisations might also operate within their own ethical framework, particularly where its purpose is societal or political change.

## The decision-making process

Whether formally or informally, all organisations will have processes for monitoring performance against objectives and also for surveying the wider environment in order to identify risks and opportunities. This monitoring will be reviewed by decision makers in the organisation, who will decide whether an intervention or change in course might be needed.

Taking a decision implies some sort of active choice. However, the absence of a decision does not mean there is no scope for choice. Whether or not it is recognised as such, reviewing information from monitoring activity and doing nothing in response is a decision to remain with the status quo.

From time to time an organisation will recognise the need to change direction in some way. This could be anything from identifying a need to adjust a portfolio to remain within risk limits, to a recognition of a need for a fresh strategic direction for the whole organisation. There will usually be some sort of default if a decision is made not to change course. This default is commonly referred to as the counter-factual, the status quo or "do nothing".

The organisation will have processes in place (once again, more or less formally) to identify alternative courses of action (including "do nothing"), analyse the likely consequences for each and present that analysis to decision makers. This process will require an assessment of whether any of the alternative courses of action are better than the status quo, and which of the alternatives is the best. Such assessment should properly have a consideration of the risk and uncertainty associated with all options.

It is often the case that there are many alternatives courses of action facing decision makers, and these usually can't be neatly defined as a small number of discrete options (e.g. Options A, B and C). This can either be because a decision is multi-dimensional in nature, or perhaps that decision makers have to make a choice along a continuum. Structuring these options for analysis and presentation to decision makers and stakeholders is an important part of effective decision making. It will be necessary to strike a balance between the number of options that can be modelled and presented and the risk of inadvertently excluding important options or combinations of options.

Once the decision has been made and the organisation's course of action has been set in motion, the consequences of the decision will then be monitored as a part of the organisation's processes. In this way the cycle of decision making starts again (see below).

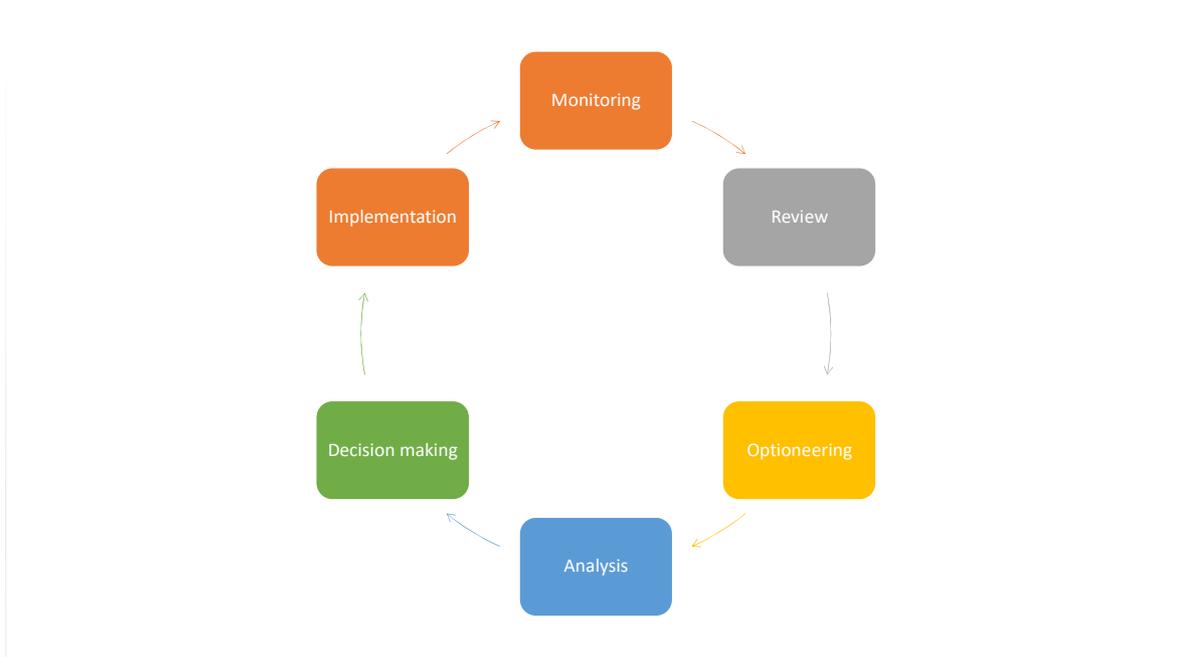

To the extent that an organisation is a formal legal entity, decisions will be made within a formal governance framework, which would define how decisions are made and by whom. This decision-making authority might by exercised directly or delegated – often down to an operational level. For example, a commodity trader will have the delegated authority to change positions in a market within agreed limits. A change in the risk limits would require a higher level of delegation. A change in risk policy might require a board level decision.

## Outcomes – public and private value

Any decision will commit the organisation to a course of action that will have consequences for future outcomes. Outcomes are those variables, quantifiable or subjective, that determine whether an organisation is furthering its objectives or purpose.

For example, a commercial organisation, the outcomes could be measured in terms of financial gain or loss, but it could also be in terms of intermediate objectives, such as market share, brand awareness or production cost.

For a non-commercial organisation, the outcomes will often be something other than financial gain or loss, but financial impact or efficiency will often be important as a means of furthering the primary objective.

A proper description of future outcomes will include a description of risk and uncertainty. The options considered will vary both in the expected outcomes and the degree of uncertainty around them, and decision makers will need to consider the best combination given the purpose of the organisation and the preferences of stakeholders.

This multi-variate mix of outcomes, quantifiable and subjective, each with its own profile of risk and uncertainty can be captured in the concept of value. The objective of an organisation is to enhance the value to the stakeholder group defined in its purpose. For a shareholder-owned commercial organisation, the appropriate measure of value would be shareholder value. For government it would be some measure of public value.

In all cases, a proper consideration of value would include qualitative as well as quantitative considerations and would also reflect the degree of risk around outcomes. For example, shareholder value could be enhanced by an improvement in a company's brand profile or reputation for social responsibility as well as enhanced financial outcomes. Value would also be enhanced by a reduction in the degree of risk around future returns.

A public sector organisation's *public* value can be enhanced not only by delivering outcomes, but also through developing its strategic vision and establishing support for this vision through the political process (its licence to operate). In this way, the risk to future delivery of benefits to citizens can be reduced.

For a commercial entity it may be appropriate to try to express all aspects of a decision in financial terms, particularly given that there can be an ultimate expression of this value in share prices or transaction values.

For public value, it may not be possible, desirable or even ethically acceptable to express all elements of a decision in financial terms. However, some means of prioritisation or trade-off needs to be considered between elements of public value, even if this is not through the medium of monetary equivalents.

## Using models to help with making decisions

Any decision needs to have regard to the possible outcomes arising from a course of action. This means the decision maker needs to have a model of how outcomes will be affected by the various alternative course of action available. This model can be used to explore the uncertainty around those outcomes.

At one extreme, a model could be a sub-conscious, conceptual construct, based on experience and beliefs about how the world works: intuition or "gut feeling". Or, at the other extreme, it could be a complex computer-based model making multi-variate forecasts many years into the future under a range of scenarios. In practice, decision makers use both formal and informal models in support of their choices.

Theories of management, finance and economics are often used to support decision making in organisations. However, these theories themselves are also forms of model, embedding a priori assumptions about the way the world works. Any modelling needs to be aware of the a priori assumptions and implicit models embedded in these theories.

Even at the monitoring stage of the decision cycle organisations will often use modelling to support decisions. For example, forecasts of expected year-end out-turn using models based on the most recent performance data might be used to assure decision-makers that the organisation is on-track to meet financial expectations. Financial organisations will often model their exposure to particular risks, ensuring their positions in the market remain within risk limits. An energy supplier will model the exposure of its portfolio of energy contracts to changes in wholesale energy prices.

Any model embeds assumptions about the way the world works and uses data from observations about the world today. These assumptions are often not well supported by evidence and are not always ethically neutral. This introduces new sources of uncertainty or bias for the decision maker. Good decision making would require an understanding of (or a sufficient degree of assurance about) the a priori assumptions embedded in the model and the uncertainty inherent in model data and functioning.

## Multiple Models in support of decision making

An organisation will use a variety of models to support its operations and decision making. These models will vary in complexity, will be more or less formal and will operate in different domains.

The use of multiple models in organisations is the norm rather than the exception. Indeed, it is not unusual for multiple models to be used to support a specific decision. Any approach to the use of models for decision making in organisations needs to address how best to manage and make use of evidence from multiple models, including how formal modelling evidence is considered alongside other forms of model, such as economic theory, non-quantifiable factors and subjective consideration.

Inconsistency between models is not necessarily a bad thing. A full description of any modelling evidence should include information about input data quality, modelling assumptions and the treatment of uncertainty. If there is full transparency, two models using different approaches could provide valuable insight that could improve the quality of a decision. However, if models are inconsistent but are unwittingly treated as equivalent, this could lead to poor decision making and adverse consequences for outcomes – indeed a common circumstance is for different parties in a debate to have their own models, with specifications not being shared, and proper comparison thus being impossible.

In some cases, models are linked, so that the outputs, or insights, from one model are embedded in a second model. This is most commonly the case in economic modelling, where models of behaviour or economic theory can be embedded in a formal computer-based model. For example, a model might implicitly assume perfect foresight on behalf of market participants by using an ex-post optimisation approach.

More specifically, it is important that the information about uncertainty and the a priori assumptions embedded in one model is not lost when outputs are transferred to another model. That is true of both conceptual models and formal models.

## Summary

In summary, all organisations use models in support of their decision making, even if they are not aware of doing so. In most, if not all cases, multiple models will be used in an organisation, which will vary in their approach and degree of formality.

An organisation wishing to improve the quality of its decision making will be need to be fully aware of these different models and what they add to the evidence base. This would require a full understanding not only of the model outputs, but also uncertainty and risk around them, whether this arises from input data, the model assumptions or inherent stochasticity.

# 3. Uncertainty in decision making

## 3.1 Background

Uncertainty is central to any discussion of the role of modelling in decision making. The entire purpose of using models to support decision making is to reduce or understand uncertainty around the possible impact of certain actions. It is to place the decision maker in a better position to make a choice.

A model typically incorporates established or imputed relationships between the projected outcomes (which are inherently unknown or uncertain) and input data or assumptions (which are usually known with greater degree of certainty). Any model outputs presented to the decision maker must include information about both projected outcomes and their uncertainty.

Our discussions have benefitted from recent initiatives in the area of uncertainty quantification, notably the recent Isaac Newton Institute programme [13], and two recently completed Engineering and Physical Sciences Research Council Networks: M2D (Models2Decisions) [14] and Cruisse [15] (each of which produced final reports and future research agendas).

Much progress was made but there is still urgency to improve the theoretical understanding in relation to established traditions, notably statistics, modern data science, applied mathematics, economics and philosophy, and the interactions between them.

At the same time, because of critical decisions to be taken in energy, transport and other infrastructure and the huge and perceived risks facing mankind, such as climate change and the risk of ecosystem collapse, understanding uncertainty in decision-making is of paramount importance.

The increasing importance of uncertainty acknowledged in official reports but not yet reflected in practice. This may be because there is a lack of understanding of how to handle it mathematically and express it in communication.

Even risk assessment, which is one of the better developed methodologies, still has a long way to go, for example in taking a thoroughly multi-objective and systemic approach. At worst, the assessments are inadequate and ignored, despite the demands of regulatory authorities.

There is an urgent need to promote a better understanding of uncertainty in decision making, including the sources of uncertainty, its relevance to the decision at hand and the means of handling it in modelling and representing it to decision makers.

## 3.2 Decisions, Models and Uncertainty

Decisions are made in the face of an uncertain future, and with imperfect knowledge about the present. An organisation chooses to pursue a particular course of action because it believes that this it is likely to result in better outcomes than the alternatives.

Properly, such a decision would take full account of the associated uncertainty and the risk to the projected outcomes. Indeed, in many cases, decisions are made specifically to manage exposure to uncertainty. For example, a decision to take out an insurance policy seeks to enhance overall value by reducing the financial downside from a catastrophic loss. In such cases, uncertainty is inherent to the decision rather than simply a risk to be considered or managed.

The whole objective of modelling as a part of decision making is to reduce the uncertainty around future outcomes. Without modelling, a decision maker would be dependent on heuristics or informal conceptual models (often referred to as "gut feel"). Modelling derives information about uncertain outcomes from elements that are known with greater certainty. For example, a simple model could be used to estimate production costs with greater certainty from knowledge about the cost of input materials, running costs and capital costs.

However, the use of a model does not eliminate uncertainty. There is uncertainty inherent in the input variables, some of which may themselves be based on model-derived forecasts. Some of input variables may be inherently stochastic, such as weather-related data. There will also be uncertainty arising from the model itself. The model will not be a perfect reflection of the reality is representing, and will generally involve approximation and compromise, particularly around the finite resolution of time and space. This is particularly true when models incorporate assumptions about human behaviour and decision making rather than simply physical systems. Most economic models fall into this category.

Viewed in this way, uncertainty isn't just an important but neglected part of modelling in support of decision making, but its whole purpose. It is impossible to determine whether a model adds value to a decision without knowing something about the uncertainty associated with its forecasts.

## 3.2 Taxonomies of uncertainty

One of the challenges in dealing with uncertainty is a certain imprecision about the use of terms, such as risk and uncertainty, which can be used interchangeably and also used to describe quite different concepts without clear distinction.

There are a number of taxonomies of uncertainty available, which come from different perspectives – some are more aimed at model-based predictions, and some are more general. Summarising these can be helpful, as they all have their role in helping think about uncertainty in specific cases and the methods which might be used to support decision making under uncertainty. Thus these taxonomies co-exist, and all might help provide some insight for any given situation. The purpose of this section is thus to provide the reader with a range of different perspectives and ways of framing decision questions.

### 3.2.1 Aleatoric versus epistemic

This commonly-seen distinction contrasts

- *Aleatoric uncertainty*: uncertainty which cannot be reduced through experiment or analysis, such as the outcome of a single coin toss. This is commonly referred to colloquially as *intrinsic randomness* or similar.
- *Epistemic uncertainty*: uncertainty which can be reduced by further analysis or experiment, such as the probability of a head if one is not sure in advance that the coin is fair.
- *Deep uncertainty*: this term is often used to classify some epistemic uncertainties, where there is little possibility of finding any kind of consensus on quantification of the uncertainty. This should not be taken to include cases where there is consensus on a broad picture but not on detail, which can be handled by sensitivity analysis to the precise quantification used.

It is important to note that epistemic uncertainty cannot necessarily be eliminated entirely – it is simply that one can learn more through additional thinking or experiment.

Also, for some ways of thinking about specific situations, the distinction may be moot. For instance, if one is interested in the outcome of a single coin toss, the examples of aleatoric and epistemic uncertainties described above combine into uncertainty about that single event. It is worth thinking about this example when one sees statements such as 'encoding expert judgement in terms of probability [i.e. a subjective Bayesian picture] does not mean real probabilities' – when it comes to thinking about the outcome of the coin toss, one needs a framework for combining different aspects of uncertainty and judgments, otherwise one just has the ability to talk about uncertainty in the outcome conditional on a given probability of a head.

### 3.2.2 The 'Cynefin' framework

Cynefin is a Welsh word meaning 'habitat', and this framework was developed by Snowden as a way of categorising in a general context degrees of uncertainty. Quoting French[1], the categories are:

- *Known*: The Realm of Scientific Knowledge. Cause and effect understood and predicable.
- *Knowable*: The Realm of Scientific Inquiry. Cause and effect can be determined with sufficient data.

---

[1] S French (2013) Cynefin, statistics and decision analysis, Journal of the Operational Research Society, 64:4, 547-561, DOI: 10.1057/jors.2012.23

- *Complex*: The Realm of Social Systems. Cause and effect may be determined after the event.
- *Chaotic*: Cause and effect not discernible.

This can be helpful both in framing the nature of a decision study, and more specifically in considering the kinds of technical modelling approaches which might be relevant.

### 3.2.3 Sources of uncertainty in computer models

Section 4.2.2 contains a useful taxonomy of sources of uncertainty in computer models developed by one of the present authors (Goldstein). In other presentations, Goldstein further points out that there is often uncertainty about what is meant by *uncertainty* and *probability*. For instance, frequently one sees that a study will claim in some sense to have "done uncertainty", whereas often there will be an issue like a conditional statement of uncertainty being made given certain strong assumptions (and consequent danger of this being misinterpreted as an unconditional statement of uncertainty based on a much broader uncertainty analysis); or there will be some important misunderstanding around the use of the word probability (in particular commonly in decision making there are relevant one-off events, and thus a frequentist understanding of probability based on the idea of repeated trials is not relevant).

Some examples of uncertainty issues from the experience of the authors, which might be helpful in thinking about Section 4.2.2, are as follows:

- Uncertainty around decision variable (i.e. factors determined by the decision maker)
    - Completeness of the option set
    - Description of the counterfactual
    - Scope for delay, subsequent changes of course, risk mitigation, exit routes
    - Does the decision affect the other supposedly independent variables (e.g. price of input materials)?
- Uncertainty around input variables
    - Absence of knowledge about input variables and their uncertainty
    - Measurement error
    - Stochastic uncertainty
    - Correlation (and auto correlation)
- Uncertainty around the model fidelity
    - Model granularity (finite resolution in time and space)
    - Stability of the rule set – from laws of physics to market regulations.
    - Uncertainly about behaviour and motivation of agents, such as other decision makers, consumers and governments.
    - Approximations and simplifications of complex realities

- Incomplete representation of outcomes, and their relationship with the various elements of the options available.
- Inadequate communication or visualisation to decision makers.
- Black swans

### 3.2.4 Uncertainty in decision support studies

The following broad classes of uncertainty in decision support studies are enumerated by French[2]:

- Uncertainty in knowledge of the external world: stochastic or aleatoric uncertainty, epistemic uncertainty, uncertainty over how other actors will behave
- Modelling and Analysis Errors: judgments in model parameters, computational uncertainty, other aspects of model-world difference
- Internal uncertainties about ourselves: ambiguity and lack of clarity (for instance in verbal expression of information), value, social and ethical uncertainty, uncertainty in what depth of modelling is sufficient

This has proved useful in framing how complex decision questions are mapped on to quantitative modelling studies.

### 3.2.5 Knightian uncertainty

Knight's discussions of the meaning of uncertainty[3] and probability are very often quoted. Among other matters, Knight refers to distinctions between

- (From Chapter VII) *a priori probability,* and statistical *probability* and *estimates*, the former referring to situations where there is some physical reason to assign probabilities (e.g. the roll of a fair die, or combinations of National Lottery numbers), and the latter referring to situations where some form of inference from numerical data or other forms of knowledge is required;
- (From Chapter VIII) *risk* versus *uncertainty*: quoting at length, "The practical difference between the two categories, *risk* and *uncertainty*, is that in the former the distribution of the outcome in a group of instances is known (either through calculation a priori or from statistics of past experience), while in the case of uncertainty this is not true, the reason being in general that it is impossible to form a group of instances, because the situation dealt with is in a high degree unique". He also calls these two concepts *measureable* versus *unmeasurable* uncertainty, and refers variously to estimates and unmeasurable uncertainty as *true uncertainty*.

---

[2] Personal communication based on unpublished work.
[3] F. Knight, Risk, Uncertainty and Profit, Riverside Press, 1921. At the time of writing, a pdf copy is available for download at https://fraser.stlouisfed.org/title/110.

We do not attempt here an exhaustive critical review of Knight's thinking, but do wish to sound a cautionary note about contemporary interpretation (or maybe misuse) of Knight's work. Knight makes important points, which stand the test of time, and it is important to note that in his book he has a particular interest in the ability to monetise risk through statistical aggregation and insurance.

It is important when referencing Knight to do so in the context of a further century of evolution of the understanding of probability and other frameworks for managing uncertainty. One particular issue is where Knight is explicitly or implicitly evoked as an authority carrying the message that for 'unmeasurable' uncertainties one cannot use probability. Any concerns over probabilities being *subjective* in this context apply equally to any other quantification of uncertainty, one then forgoes use of the apparatus of probability theory for managing and combining uncertainties, and decision analysis based on interval approaches to quantification can provide results which are extremely sensitive to the precise details of the interval limits.

We are not saying in this section that probability is necessarily always the answer, but rather that decision support approaches should be chosen for sound reasons, and that often reasons for rejecting probabilistic quantification of uncertainty are unsound.

## 3.3 Tools and methods for managing uncertainty in decision making

As previously discussed, a typical decision will be based on seeking to maximise value, however defined, across a range of qualitative and subjective outcomes. In many cases, less certain outcomes will be less valuable, and often significantly so. This is particularly the case where resilience or security is a major consideration.

Assume there is a clear understanding of the options available to the decision maker and these are well structured.

### 3.3.1 Models and system integration

Much systems work involves individual computer models talking to each other in various ways. So, for example, we might have a chain of models in which some of the outputs from some models are some of form some of the inputs from the next models. More complicated could a trees, network, ensembles etc. of models. This is understood in advanced scientific or engineering. For example, "probabilistic engineering" and "robust design" are gaining ground in the automotive and aeronautical sectors.

The way to assess uncertainty for the combined system, over the whole range of input choices, is as follows. Uncertainties typically build up as we move along a chain of models, up the tree of models to the leaves, or outward to the extremities of the full systems. But internally there are important bottlenecks and critical points of sensitivity. For example, in epidemiology infection rates are critical. Special components or subsystems may require special attention. Different parts of the system may be populated with different data and different types of data. Typically there will be mathematical assumptions based on rules, theory or laws of some kind and other parts based, and calibrated by, empirical data. Thus the models for and engineering system will have material properties as parameters in the system, which are known but are based on separate laboratory testing. Put crudely different components or subsystems made need special experiments be they computer experiments (simulations) or physical experiments.

Surrounding any model is a no-man's land of factors which may bias the model but are not included in the model. Some of these are known and some not. These could be physical, but also they may be the effect of human intervention from hands-on human factors, as in air-traffic control, to longer range policy decisions, such as earlier release of prisoners.

In many discussions and seminars of these issues a major objective emerges: how to quantify uncertainty of the whole systems. This can be refined to (i) how to capture the uncertainty in the component models (ii) how to integrate this uncertainty into uncertainty evaluation of the whole system.

Here are additional issues (see also the points in the Introduction).

### 3.3.2 Fidelity

Many models can be designed/run at different levels of fidelity or resolution. Thus is a finite elements method one can use different mesh size, in a climate model the resolution (size of grid box) and in social surveys one may have different sample sizes or different levels of stratification. High resolution models are typically more expensive to build and to run. There is a very active research on combining models of different fidelity, balancing cost and accuracy.

In a network of models the component models have, and probably should have different levels of fidelity. Spreadsheet type models are the backbone of accounting, but they are at coarser resolution than, e.g., an engineering model. Different classes of model need to communicate, for instance the discipline of cost accounting uses both engineering and accounting models.

A theme running through this report is that models should reflect the purposes they are put to. A more refined version of this would be that the level of fidelity should be tuned to use – for instance in some circumstances a low fidelity, fast-to-run, easy-to-understand model may be enough. [16]

### 3.3.3 Control, options, adaption.

In Economics Nobel prizes have been awarded for option pricing which employs advanced methods from stochastic processes. Indeed the term stochastic process typically implies one or more random variables indexed by time (of course, space too, sometimes). Time is the key variable in control and without it there is no concept of the feedback on which control is based.

So, although many models start out as static, time will be forced upon them because, as made clear already, they are used to forecast the future or to play what-if games with the future. Decisions themselves take place over time. Even a supposedly one-off decision can come back to haunt the decision-maker. There may be re-thinks, redesign, updates, servicing, monitoring, or progress reports. Bridges crack, prisoners reoffend, time-tables fail, blackouts occur, demand drops.

The best high-level mathematical models to cover these features are some kind of decision tree with options at certain points (vertices, nodes) and analysis tracks the implication in terms of cost and benefits of different paths along this tree. Then, if one is lucky one may be able to optimise the route. So, ideally, our network of models covers multiple times, and has decision points labelled, the modelling needs along the decision paths understood and the data needs assessed. [17].

### 3.3.4 Scenarios.

Sensitivity analysis, stress testing and similar fields are fairly well developed. The "softer" area of scenario analysis is one way of handling the possible biases referred to above. They will also affect different parts of a modelling exercise. There may typically be a preferred scenario (with uncertainty) and then uncertainty may be predicated on different scenarios.

In a recent workshop several aspects of the use of scenarios in decision-making were highlighted [18]: (i) scenarios to capture what might happen versus what we can make happen; (ii) scenarios for short- or long-term decision making; (iii) how scenarios can fill the space of uncertainty, with or without the use of probability; (iv) the need to critique existing methods such as those based on "regret"; (v) the design of scenarios as a creative tool to stimulate open minded approaches to the future; (vi) the relation to expert judgement.

### 3.3.5 Expert judgment, elicitation and cognitive bias

A fundamental issue is how to combine expert judgements with modelling. How to combine expert judgement is a fundamental. Bayesian methods have a clear mathematical and philosophical foundation for making them, to date, the natural first choice [19].

Structured elicitation can be used to improve over simple "guessed values" for unknown values of important variables. Probabilities can be elicited, whether formal prior distributions, or as input to some kind of simulation based sensitivity. In interval-based and worse-case analysis maximum and minimum values can be elicited.

Behavioural economics has given us formal ways to handle the cognitive biases encountered in elicitation. Such biases have notoriously been responsible for time and cost overruns in recent UK capital projects. A particular field within expert elicitation is thus growing under the title of "de-biasing" [20].

So to our vision of multiple models in the service of decision makers we need to add what may be a new hybrid methods crafted from the best of Bayesian elicitation, sensitivity analysis and the practical ends of behavioural economics.

The previous decisions are in many cases the most important variables in the next decision. We are not moving into a world where algorithms can make decisions divorced from humans. The driverless vehicle is designed by a human being. Models have many components where is insufficient data to determine the precise current value of a parameter, let alone the future value. Climate change depends radically on decision by governments, companies and citizens. Climate models do not make the decisions. There is a larger picture in which human action is not just the most important variable, but also the biggest source of uncertainty.

### 3.3.6 Domains of modelling

There is a need for a taxonomy of models in an organisation, based on the different domains in which they operate. Typically, one finds nested domains, with each level being an emergent property of complexity in the level below. What this means in practice is that the level of complexity at the lower level is so great that it is no longer possible to understand the higher level system in terms of the lower level components. It also means that the nature of modelling and uncertainty in one domain is sometimes very different from those in other domains

Some broad categories include:

*Physical, engineering, technological*

*Climate, geophysical*

*Biological, medical*

*Behavioural, cognitive, psychological*

*Epidemiological, ecological*

*Economic, financial, accounting*

*Organisational, management, logistical*

*Sociological, cultural*

*Historical, political*

In model hierarchies each level is an emergent property of complex adaptive systems at the previous level. The uncertainty relates to the system function at that level, but also makes a priori assumptions about the functioning at the lower levels. For example, economic models make assumptions about behavioural characteristics of individuals. Biological models are dependent on an understanding of chemistry. However, we do not understand life sciences primarily in terms of chemical or mechanical processes, but rather use models and concepts about the functioning of biological systems.

Multiple models is thus in part the use of multiple models in the same domain, but also potentially about being clear about the implicit models of systems at lower domains. For example, a model of the electricity market needs a sound foundation based on the physics and engineering of electricity systems.

# 4. Methodology for handling uncertainty

The other Chapters of this paper present material in a form broadly accessible to those involved in decision support. This Chapter presents a more technical discussion of methodology options for handling uncertainty. Section 4.1 provides a broad survey of issues, and then Sections 4.2 onward provides a detailed technical description of technical matters, where relevant using a Bayesian framework. Thus while we encourage readers to engage with Chapter 4, it is not critical to an understanding of the rest of the paper.

## 4.1 Introduction

There have been many attempts to understand and formulate the concepts of uncertainty, from early work by Knight to work by various bodies charged with decision support issues regarding risks facing mankind such as the work of the Intergovernmental Panel on Climate Change (IPCC). Distinguished thinkers have contributed such as Keynes' ideas regarding "weights" and "specificity" and Carnap's theory of evidence. This is a rich heritage which falls into various categories. At the forefront must be solving the "problem of induction" posed by Hume with the exposure of the tension between subjective belief and raw data. To give the conundrum a modern flavour we can ask: how exactly do we combine expert judgement with data?

It can be generally agreed that Bayesian methods in which prior belief, typically about parameters in a model, expressed via probabilities, is combined with data, ("sampling" to use the classical term) to produce posterior beliefs again represented as probabilities, is probably the best method in town. It has and continues to have champions.

The history is that there roughly a split between the rather hard-nosed statistical decision theory of Abraham Wald and the softer subjective theory of di Finetti. The former theories show via the "complete class theorems" that many methods of classical statistics are "admissible", that is, give a utility cannot be bettered in a Pareto sense if and only if they are Bayes rules. Put informally, even if you don't want to be a Bayesian you are behaving like a Bayesians with respect to some prior distribution. The theories of di Finetti appeal to gambling to show that a choice of prior distribution and the Bayes rules that follow are coherent in a well-defined way. Whether or not the prior distribution is justified in some absolute sense the Bayes machinery leads to consistent decision making.

A notable contributor to the early period on Bayesian theory was Savage and the utility-based methods under the catch all of "decision theory" have dominated many fields, particularly economics and financial mathematics. It is a remarkable historical fact that that game theory (Von Neuman and Morgensten), and related areas of optimization, such as Linear Programming, arrived at about the same time. Wald was talking about minimax theory at the same time as those authors: a minimax rule can be a Bayes rule with respect to a "least favourable" prior distribution.

A principal issue today is whether these theories are sufficient to capture all we need to understand the relation between uncertainty and making actual decisions. One indication of the urgency of the issue is the many workshops and conferences with titles like 'decision making under uncertainty" and the re-emergence of "decision support systems", including the current project. Another indication is the large effort in trying to fuse statistical methods and applied mathematics, complicated by the fact the applied mathematics had already coined the term "Uncertainty Quantification" to capture highly technical areas such as stochastic finite element methods, and some aspects of sensitivity analysis and inverse problems.

The easier "extrinsic" methods propagate variation from inputs to outputs, a method common to nearly all fields and well understood by the term Monte Carlo methods. They are used, for example is stress testing financial instruments. The "harder" methods embed the stochastics right inside the differential equation solver to give "stochastic finite element" methods. A caricature version of the split between applied mathematics and statistics is that statisticians don't know any differential equations or approximation theory, and applied mathematicians don't know any statistics. Emulation, discussed in the next sections, has gone some way to bridging the gap.

But the jury is out and may remain out for a long time attempting to pin down a theory of, or even a good guide, to uncertainty which is useful in practice. One can perhaps see the outline of such a guide as some kind of "tree of uncertainty" with increasing levels of liberality as we go up the tree. At the bottom would be the rich theories of mathematical statistics, utility and belief; a zone in which volumes of data and good elicitation of expert judgement would drive out philosophical debates about induction. In a way this is the promise of the big-data-machine-learning-AI era that fast data collection and fast computers have thrust us into.

At the top of the tree is Knightian uncertainty, a rarefied zone, where we may not know what we don't know and where rare catastrophic events can occur with probabilities we don't know and effects we hardly know; where we feel we have to be "precautionary", but we cannot afford to be (see also discussion in 3.2.5 on appropriate reference to Knight). A sort of "anxiety zone".

Between the top and bottom of the tree lies classical statistics: statistical method which trudge along without prior distribution, but using classical confidence regions. The latter has to be understood (although students find it hard) as a random set theory: the confidence region is a random set and we declare that it "covers" the unknown parameter with a known probability (or a lower bound). This in itself is a gamble, but a different sort of gamble to those embodied in Bayesian theory. It is an objective gamble given by the randomness in the data (sample generation).

At about the same level in our uncertainty tree are two other areas: Dempster-Shafer believe theory and Fuzzy set theory. Without going into the details one can say that they are also set-based theories whose axioms are close in spirit to the "coverage" theory of classical statistics. If we are to define more clearly the levels of the tree it is likely that we will draw from the catalogue of axioms from these and related fields. Simplistically we climb the tree by weakening the axioms.

If these areas are to compete with the subjective Bayes methods they must develop fast computation. For example, there is much scope for introducing fast computational geometry into the construction of confidence regions; sets are hard to deal with computationally. There is already much geometric thinking to exploit contained in the duality between confidence regions and hypothesis testing: "covering" a value of parameter value is equivalent to accepting that value as a "null" hypothesis.

Any success with the venture will allow us to ascertain more clearly which other activities should be used at which level of uncertainty: data collection, optimization, elicitation, modelling, and so on.

Despite the lack of integration among all the subjects mentioned above some areas have emerged to on which there is general agreement.

1. Objective Bayes. With special choice of prior distributions the coverage properties behaviour of Bayes rules can be close to that of classical statistical rules.
2. Bias. All methodologies can be concerned with the elimination of bias from known or only vaguely known sources. Randomization, as in Randomized Clinical Trials (RCT) is spreading from medical into several other areas as a cheap way of balancing against bias, so that causal models are not corrupted. Understanding sources of bias is critical.
3. Mean-risk, mean-variance. A common feature of several areas is the need to optimise an output or keep it on target while at the same time minimising variability. The areas include Portfolio Theory (Markovitz), stochastic control, robust engineering design, variance reduction such as in the use of antithetic variable and (recently) multi-scale simulation. Put simply, a utility function should express variability as well as mean behaviour. A related area is reliability where we want to keep the output inside some "safe domain".
4. Entropy and information. Increasing belief is closely related to increasing information. And conversely, low information (= high entropy) can express ignorance. Information theory is among the most well-established sister areas of decision theory and a cornerstone of communication theory. From choosing maximum entropy priors in objective Bayes methods to its use in optimal data collection it is a good prediction that theories of uncertainty will benefit from theories of information.

## 4.2 Technical issues in quantifying and managing uncertainty

In the following sections, we shall discuss the technical issues involved in the treatment of uncertainty in collections of models. These are as follows.

A. A discussion as to how uncertainty is treated in general in organisations, and how such treatment might be improved.
B. A discussion of the role of models in this treatment of uncertainty.
C. A discussion of the special features which arise when dealing with multiple models.
D. A discussion of how this treatment of uncertainty relates to decision support within organisations.

In order to give a unified treatment of these topics, we will present the material from a Bayesian viewpoint, in which each uncertainty statement represents the judgement of a relevant expert, or the consensus judgement of a group of experts. There are various other notions of uncertainty which may be helpful and insightful in dealing with individual portions of the analysis, but we do not know of any other formulation which could be used to present an integrated and principled treatment of all of the issues that must be addressed when considering uncertainty quantification and decision support within in a general model based context. These issues arise because of the many sources of uncertainty that must be analysed when dealing with large scale real world problems, as all of these sources must be quantified and integrated into an overall collection of uncertainty judgements which are appropriate and sufficient to support decision making for such complex problems.

### 4.2.1 The general treatment of uncertainty.

Uncertainty is not well handled in most organisations. This is for two main reasons.

(i) Uncertainty is poorly quantified for most problems, both in terms of careful and detailed identification of outcomes of interest and also in considering the consequences of such outcomes (e.g. the level of power failures resulting from a particular amount of investment in power systems, and the likely consequences of such failures).
(ii) The way in which uncertainty assessments feed into decision making is usually inadequate. There is little exploration or understanding of the real risks involved in the competing decision choices, still less any attempt to optimise the inherently sequential problem of multi-attribute decision making which the organisation faces.

In practice, (i) and (ii) are related. Because uncertainty assessments are not used seriously for decision making, there is little motivation or perceived value, and hence little allocation of resource, for creating such assessments carefully. Because the assessments are not carefully made, there is little benefit to be gained from actually using them to direct decision making.

Improvement starts with a more careful and consistent approach to the treatment of uncertainty in decision making. At the fundamental level, a process of improvement would be as follows:

(i) to be explicit as to the main sources of uncertainty governing the outcome of the decision problem;
(ii) to make an honest attempt to quantify each such uncertainty;
(iii) to compare critically the risk profiles (possible outcomes with associated probabilities) of competing decisions as a basis for choosing a safe and reliable decision;
(iv) to assess the value of different potential sources of additional information (by assessing the likely reduction in uncertainty expected from each such source, and hence the expected change in the risk profile for each decision) and hence to ensure that the decision process is well-informed;
(v) to construct a clear audit trail by which the rationale for each aspect of the analysis can be recorded and can later be assessed against actual decision outcomes to validate and improve the decision-making process in the organisation.

With such systems in place, it is then possible to explore the value of careful uncertainty quantification for achieving better (safer, more cost-effective) outcomes. How such improvements in uncertainty assessment are made is problem specific. However, there are common features underpinning most applications. In particular, our focus is on model based uncertainty, which we now discuss.

### 4.2.2 Model uncertainty
Models provide powerful ways of exploring competing decisions. As such, they are an essential tool in dealing with complex problems. However, this power is potentially misleading if we overlook the uncertainty associated with every statement that is made within a modelling framework.

A careful treatment of the uncertainty arising in model based analyses requires considerations of the following aspects of uncertainty.

(i) PARAMETRIC UNCERTAINTY (each model requires a, typically high dimensional, parametric input specification, whose choice is unknown).
(ii) CONDITION UNCERTAINTY (uncertainty as to boundary conditions, initial conditions, and forcing functions, such as future climate, future energy demand, etc).

(iii) FUNCTIONAL UNCERTAINTY (model evaluations take a long time, so the model output is, in effect, unknown for every input choice where a model evaluation has not been made).

(iv) STOCHASTIC UNCERTAINTY (either the model has stochastic components, with immediate implications for uncertainty analysis or it should have such components and stochastic features of the system have been suppressed).

(v) SOLUTION UNCERTAINTY (as the system equations can only be solved to some practical level of approximation within the computer implementation of the model).

(vi) STRUCTURAL UNCERTAINTY (the model is not the same as the real physical system, so that good system outcomes within the model do not ensure good real world outcomes).

(vii) MEASUREMENT UNCERTAINTY (as the model is calibrated against system data all of which is measured with error).

(viii) MULTI-MODEL UNCERTAINTY (usually we have not one but many models related to the physical system of interest)

(ix) DECISION UNCERTAINTY (to use the model to influence real world outcomes, we need to relate things in the world that we can influence to inputs to the simulator and through outputs to actual real world impacts. These links are uncertain.)

Because of the complexity in addressing all of these sources of uncertainty, in practice they are largely ignored, and instead, typically, some form of scenario analysis is performed. This involves making a few, hopefully well-chosen, evaluations of the model, and testing out certain competing decision choices within each scenario, to gain insight as to the comparative effectiveness of each choice.

While a careful selection of scenarios can provide an insightful and informative starting point for an uncertainty analysis, none of the scenarios will actually occur in the real world, quite apart from issues of structural discrepancy (the model is not the same as the real world) which are similarly ignored. Therefore, what such an analysis does is to pass all responsibility for applying the analysis in the real world to the decision maker, who knows very little of the detail about how the model works, and still less has the ability to conduct a reasoned real world uncertainty analysis on the basis of the scenario narratives.

For each of the sources of uncertainty listed above, there is a general methodology for uncertainty quantification. Good practice is to proceed as follows:

(i) each source of uncertainty should be assessed qualitatively, to understand its potential importance for the decisions under consideration;

(ii) the most important and accessible sources should be quantified, using the general methodology of uncertainty quantification;

(iii) an order of magnitude assessment should be made for the effect of all of the remaining sources, based, for example, on expert elicitation.

The care with which each part of the analysis is carried out depends on various factors, in particular

(i) the importance of the problem under investigation. This is obvious - though often we observe the opposite, namely that much more careful treatment is given for relatively small sub-problems, because they seem to offer themselves more easily to rigorous quantitative analysis, than is given for the large problems which are of primary interest.
(ii) the resource available (in personnel and time). The available resource is largely a management problem (the personnel available is an obvious management decision; the time available, while partly governed by real world constraints, is mainly determined by the care and foresight taken in preparing for the analysis well in advance).
(iii) the commitment of the organisation to good decision making. Again, this is obvious in principle, but too often the argument is made that we cannot afford to spend the resource now for analysis which will only give nebulous benefit at some nebulous time in the future. These benefits are only nebulous because of the bad practices of the organisation, as discussed above.

A full account of the treatment of uncertainty involves an extensive level of technical detail. However, much of the methodology can be summarised into two basic principles, which we describe as they are of fundamental importance, they lead to basic recommendations for good practice in model building, and they form the basis of the treatment of uncertainty in collections of models.

These principles relate to model emulation and structural discrepancy as we shall now describe.

### 4.2.3 Model emulation

When a simulator is too high dimensional and computationally expensive to allow us to carry out all of the evaluations that we would like to make in order to explore the range of behaviour of the model over all plausible input sets, then we should build what is termed an emulator for the model. This is a statistical model for some carefully chosen subset of the model outputs (or functions of these - for example the maximum output value over a certain time period).

The emulator is both a proxy model which gives, very quickly, an approximate assessment of the likely output of the model at any choice of inputs and an uncertainty representation expressing our uncertainty about the difference between true value of the model output and the value of the proxy model. Therefore, everything that we would like to do with the original model, but can't because it is too computationally expensive, we can do instead with the proxy model. Of course, this will be an approximation, but the degree of trust in this approximation can be assessed through the associated uncertainty of the emulator.

There are many ways to construct emulators. Often, after appropriate transformations, they are constructed as the sum of

(i) a global statistical model (for example a regression model in some polynomial functions of the inputs) expressing those aspects of the computer model surface about which we can learn from the whole collection of model evaluations that we have made

(ii) a local model for the residual difference between the true model values and the global statistical model, often expressed as a stochastic process, expressing aspects of the computer surface at any given input value about which we can only learn by making model evaluations near the given input choice.

For example, it is common to model the residual function as a Gaussian process, which denotes a function whose output has a multivariate normal distribution for any set of input choices, with a covariance between the function values for any two choices of input depending only on the distance between the two input values.

There are many ways to construct emulators. Typically, they are constructed from a carefully chosen collection of model evaluations, using standard methods for statistical model fitting (least squares, maximum likelihood, Bayes, etc.), combined with expert judgements, and supported by careful diagnostics for emulator validation.

If the model is too slow and high dimensional to allow a sufficient number of model evaluations for fitting the emulator, then we typically employ methods of multi-level emulation. This means that we develop a series of faster versions of the model (by adjusting the time step, the grid size, the solution method, the underlying system operations, etc). Then, by making many evaluations of the faster versions, identifying qualitatively the forms for the global and local models, then updating the forms given some evaluations of the full model, we are able to develop an effective emulator, even for very slow models.

In summary

(i) emulators for complex computer models should normally be constructed to overcome the difficulty of making large numbers of evaluations of the model

(ii) good modelling practice is to build models which make it easy to carry out multi-level emulation, by considering, when the simulator for the model is being built, which types of simplification should be easy to exploit in order for the emulator of the simulator to be constructed.

### 4.2.4 Structural discrepancy assessment

A computer model is a description of the way in which system properties (the inputs to the model) affect system behaviour (the outputs of the model).

This description involves two basic types of simplification.

(i) We approximate the properties of the system (as these properties are too complicated to describe fully, and anyway we don't know them fully).
(ii) We approximate the rules for finding system behaviour given system properties (because of necessary mathematical and numerical simplifications, and because we do not fully understand the relationships which govern the process).

Neither of these approximations invalidates the modelling process. Problems only arise when these simplifications are forgotten and the analysis of the model is confused with the corresponding analysis for the physical system itself, which, unfortunately, happens most of the time.

Structural discrepancy refers to the difference between the model output at the appropriate choice of model inputs and the actual value of the real-world quantities that the model purports to represent. The uncertainty associated with structural discrepancy is a central part of the problem analysis.

We may distinguish two types of model discrepancy.

### Internal discrepancy

This refers to any aspect of structural discrepancy which we can assess by direct experiments on the computer simulator. For example,

(i) we may vary parameters/forcing functions held fixed in the standard analysis,
(ii) we may allow parameters to vary over time/space.
(iii) we may add random noise to the state vector which the model propagates

We assess internal discrepancy as follows:

(i) for certain input choices, we may carry out detailed experiments, based on many choices for the internal discrepancy inputs, to determine discrepancy uncertainty, typically the variance of the internal discrepancy for an individual output and the correlation structure of discrepancy across outputs,
(ii) using emulation, we extend the internal discrepancy assessment over the input space, so that we have a full discrepancy assessment between the model and the real world physical system for all choices of input parameters.

### External discrepancy

This arises from the inherent limitations of the modelling process embodied in the simulator. It is determined by a combination of expert judgements and statistical estimation.

The simplest way to incorporate external discrepancy is to add an extra component of uncertainty to the simulator outputs. For example we may introduce, say, 10% additional error to account for structural discrepancy. This, though simple, is much better than ignoring external discrepancy completely. If we have extensive sources of data for calibration purposes, then we may adjust this judgement in light of the overall quality of the fit to data that we achieve at the best choices of input values.

Better is to consider what we know about the limitations of the model, and build a probabilistic representation of additional features of the relationship between system properties and behaviour. This notion builds on the idea that a model expresses judgements about the way in which system properties influence system behaviour, and that the information encoded into our model is informative for real world behaviour precisely because it is informative for such relationships. Typically, the result of this analysis will be expressed as an emulator for the more detailed model.

In summary,

(i) structural discrepancy should be recognised and carefully quantified
(ii) the division into internal and external analyses is helpful and informative and should be made explicit
(iii) good modelling practice is to build model simulators which make it easy to carry out internal discrepancy analyses, by considering, when the simulator is being built, how to explore ways of testing the effect of the various simplifying assumptions on which the model depends.

### 4.2.5 Model based uncertainty analysis

When the model has been emulated and structural discrepancy has been quantified, then we may use the model to carry out the various tasks for which it is intended. While these are typically problem specific, they will usually include

(i) model calibration or history matching against real world system data. This is both intended to check that the model is capable of reproducing historical observations, up to structural discrepancy and observational error, and to identify the collection of all possible input choices for the model which are consistent with such real world history;
(ii) system forecasts follow from emulation of future model outcomes over the space of input choices which are compatible with historical data, in combination with the assessment of structural discrepancy for the future outcomes, to make realistic assessments of real-world uncertainty for the forecasts and to correct for model biases;
(iii) the system forecasts usually depend on decision choices which have representation within the model. As we can forecast, with uncertainty, the outcomes under each decision choice, we can therefore compare choices and pick good decisions (i.e. having good real-world outcomes with small uncertainty).

It requires a certain amount of technical expertise to carry out all of the above. This is not surprising - we are asking technically challenging question of complex physical systems. However, if we consider such analyses to be important, then it follows that such resource should be developed. This is both in terms of trained personnel (both specialists and people with enough knowledge to carry out the more routine aspects of the analysis, while recognising when such simplified analysis is not sufficient for the task) and also developing software and exemplars to guide and support such analysis.

## 4.3 Uncertainty in collections of models and systems integration

Much systems work involves individual models talking to each other. While each system has problem specific features, there is a basic algorithm for systems integration which we now describe, based on the concepts that we have already discussed.

Suppose that we have two models where model one has outputs which, in principle, form part of the input to model two. We build emulators for the two models and assess structural discrepancy for each.

The way to assess uncertainty for the combined system, over the whole range of input choices, is as follows.

(i) Choose a value for the input structure to model 1 (this may be a fixed choice, for example a decision, whose effect we wish to explore, or a random choice over our uncertainty for an uncontrolled input, or a mixture of both)
(ii) Make a random draw from the emulator for model 1 evaluated at the input choice in step (i),
(iii) Make a random draw from the structural discrepancy specification for model 1. The distribution for structural discrepancy that we draw from may, or may not, depend on the choice of input at step (i) (for example, in the suggested method for assessing internal discrepancy we suggested a simple method for linking input choice to structural discrepancy variance).
(iv) add the values generated in steps (ii) and (iii) - this sum is now a simulation for the random input into model 2
(v) make a random draw from the emulator for model 2 evaluated at the input choice in step (iv), augmented by any other inputs (such as decision choices) that the model requires
(vi) make a random draw from the structural discrepancy specification for model 2
(vii) add the values generated in steps (v) and (vi) - this sum is now a simulation for the real world output of the combination of the two models, based on any decision choices that we have introduced.
(viii) Generating individual samples may be sufficient for purpose. Otherwise, we can treat the combined system as a single stochastic model, and emulate it, for example, by emulating the mean and variance for each input choice.

This approach is modular. We can emulate and assess structural discrepancy over each of an inter-related family of models separately, then combine all of the specifications to carry out the composite uncertainty analysis over any collection of subsystems of interest.

This approach supposes that structural discrepancy is independent across models. A more careful treatment would be to construct an influence diagram to represent this collection of uncertainty structures and their relationships. An influence diagram is a graphical representation of the conditional dependencies between all of the uncertain quantities and their relationship with all of the decision choices and outcomes within a problem. So, in our case, the nodes of the influence diagram would represent the various sub-models of the system, and the links would express the information that passed between them. The above algorithm describes the simplest such diagram - model 1 is a node which takes inputs as one node and decisions as another, and gives outputs as a node, which combines with the node expressing structural discrepancy for model 1 and a further decision node to provide a further node which expresses the input to model 2 which is a further node which leads to an output node which combines with the independent node expressing structural discrepancy for model 2 to give real world outputs for the whole system. However, more detailed influence diagrams are possible and can form the basis for more sophisticated approaches to systems integration.

## 4.4 Decision making under uncertainty

We now discus how uncertainty quantification feeds into decision making. In this section we discuss general principles of decision making under uncertainty, and in the following section how this relates to uncertainties derived from system models.

The general framework into which most decisions under uncertainty may be fitted is as follows:

> STEP 1 Specify the problem. This means identifying the possible decisions and the possible consequences, which we term rewards.
>
> STEP 2 Specify the uncertainties relating to each reward, given each decision. [This is essentially the question that all of the preceding discussion of uncertainty addresses.]
>
> STEP 3 Value the consequences. There are many ways to do this, but for simplicity we focus on the natural way to do this for rational decision making, namely to make a utility specification.

This is specified as follows. We start off with a collection of rewards, $R$. Choosing a decision gives an uncertain reward (i.e. a random pick from the reward set according to the uncertainty specification that we have made). So, choosing a decision, $d$, is equivalent to choosing a gamble, $g_d$, over rewards $R$.

A utility function gives a numerical value to each such gamble, $U(\cdot)$, satisfying two basic properties:

[U1] If you prefer gamble $g_1$, to gamble $g_2$, then you must assign $U(g_1) > U(g_2)$, so U1 says that utilities agree with preferences.

[U2] For any gamble $g$, we must have $U(g) = E(U(g))$. ($E(U(g))$ is the expectation of the utility of the gamble, for example if $g$ is a gamble which gives reward $r_1$, with probability 0.6 and $r_2$ with probability 0.4, then $E(U(g)) = 0.6U(r\_1) + 0.4U(r\_2)$.) So, U2 says that expected utility equals actual utility.

Putting [U1] and [U2] together gives us the basic solution to any decision problem, as follows.

STEP 4 If you have specified a probability distribution and a utility function over the rewards, for each decision, then you should choose the decision with the highest expected utility.

This is because the highest expected utility equals the highest actual utility from [U2] and the highest actual utility corresponds to the most preferred gamble, from [U1]. So, if you can specify probabilities and utilities over the outcome set, then the decision problem is solved.

Can you always specify a utility function over rewards? Yes. A fundamental theorem of decision theory states that under a (simple, weak and very reasonable) collection of assumptions over your preferences, you may always construct an (essentially unique) utility function over any reward set (there are different versions of this theorem proved under different sets of assumptions – the proofs proceed by constructing a utility function in a manner which can form the basis for doing so in practical circumstances). Therefore, you can always follow the above procedure and this will always identify your preferred decision.

There are simple constructive procedures to support your specification of a utility function, and if there are a variety of attributes of the decision (costs, benefits, etc.) there are natural ways in which to incorporate them into a multi-attribute utility function, or even a multi-attribute utility hierarchy (for example, costs for a medical decision might be assessed as financial and public health related, which might be further decomposed, and so forth).

Different people may specify different utility functions over the reward set so, in any decision problem, we need to identify the different stake holders and their corresponding utility functions and aim for decision choices which are good for each stakeholder.

The utility function encodes your relative preferences over rewards. It also encodes your attitude to risk. For example, given the choice between

(A) a gamble paying nothing and

(B) a gamble giving a £20 million gain with probability 0.1 and a £1 million loss with probability 0.9

then most people would prefer gamble (A) even though it has payoff zero, while the expected payoff from gamble (B) is £1.1 million.

This shows that money is not, itself, a utility function (for most people) so that maximising expected payoff is not, usually, the best way to proceed. Most people are risk averse and this is represented by, for example, treating the utility of a money amount as some concave function of money (for example $U(£x) = \log(x + c)$, for some constant $c$ which flattens the curve).

Our reasons for recommending this approach to decision making under uncertainty are as follows.

(A) Unlike all of the other approaches that are usually recommended, this method has a sound logical basis. If you follow the method, then you will identify the best decision (for you).

(B) The method requires and enforces clarity. You must explicitly state your probabilities and utilities, so that these judgements are open to scrutiny and therefore to criticism and improvement.

(C) Because different individuals may make different choices, this approach gives a natural way to identify the extent to which decisions may be found which are good for all of the stakeholders in the decision process (by comparing good decisions under each stakeholders' utility assessment) and what are the irreconcilable differences which must be confronted by negotiation.

(D) Most complex decision choices are sequential (we make choice 1, see what happens then make choice 2, and so forth). Our method provides a natural way to optimise the joint collection of choices over time (based on backward dynamic programming, for example the type of calculation that is used to solve decision trees).

(E) Because good decision analysis is both time and skill intensive for large and complex problems, the methods that we have described should be thought of as providing a powerful collection of decision support tools rather than aiming to identify a single 'best' decision. In practice, we identify not a single good decision but rather a collection of decisions which have good expected outcomes, and use the various tools of decision analysis to identify

(i) which features of the specification are most influential for guiding decision choice, and therefore, which require most care and attention
(ii) what is the value of extra information in improving the decision process and therefore how do we direct and value additional data gathering.
(iii) what are the basic trade-offs involved in our decision choices - for example, what is the tradeoff between investment cost and positive outcomes for our problem (more technically, we will typically need to identify the Pareto boundary between the most important tradeoffs).

(iv) based on the above, which decisions give the most efficient and robust performance across the range of tradeoffs and specifications that are of basic concern.

## 4.5 Decision support with models

To apply the above ideas in the context of model uncertainty, we proceed as follows. We specify decision choices d (a vector of choices, possibly over time) outcomes $f(x, d)$ where

$f$ is a computer model (possibly combining several individual models)

$x$ is the other features (for example model parameters) that we need to specify in order for the model to evaluate the outcome of decision $d$

$f(x, d)$ is a vector of the different consequences of the decision, for example $f_c(x, d)$ might represent overall costs and $f_b(x, d)$ might represent overall benefits.

We will not usually be able to evaluate $f(x, d)$ for all values of $x$ and $d$ (as $f$ is too expensive to evaluate everywhere) so we will assess $f$ using an emulator, returning, for example, an expectation and a variance for the function output for all $x, d$.

$f(x, d)$ expresses the function output not the real world outcome, so we need to add model discrepancy to create the random vector $y(x, d)$ of real world consequences of the decision choices.

The outcomes are best expressed in units of utility. In this scale, it is correct to choose the decision which maximises the expected value of the outcome.

Suppose that we are searching for the optimal decision choices $d$. We place a probability distribution over inputs $x$, supposing that we have used any relevant historical observations to eliminate those values of (sub-vectors of) $x$ which lead to historical model outputs which are inconsistent with observed history. (There is an efficient methodology to identify and eliminate such inputs, termed history matching.)

For any decision, $d$, we may assess the utility of this choice by assessing the value of $U(d) = E(y(x, d))$ where the expectation is taken over $x$ given $d$ and $y$ given $x$ and $d$. Typically, this calculation will involve the use of the emulator for $f(x, d)$.

From a decision support viewpoint, we aim

(i) to identify the highest utility, $U^*$, that we may achieve, and
(ii) to identify the collection of decision choices which achieve outcomes close to $U^*$.

so our aim is to identify the maximum of the utility function and to invert it in a region around the maximum.

We may then consider the collection of good decisions by comparing their risk profiles. The risk profile of a decision is the collection of outcomes that may result from the decision, with their associated probabilities.

In general, we look for robust decision choices, i.e. choices for which the chance of bad outcomes is small and the main features of the good outcome do not depend too heavily on the precise assumptions of our models. For example, if two decision choices achieve, on average, roughly the same outcome, but for the first choice, the model discrepancy is assessed to be substantially greater than for the second, then we would typically prefer the second decision choice.

Assessing the value of $U^*$ may be a difficult optimisation problem which relies on our ability to construct reliable emulators for all relevant function outputs. Therefore, we typically optimise in stages as follows.

For each decision, we have uncertainty about the expected utility of our decision arising from uncertainty in the emulator. However, for any pair of decisions, if the upper bound on uncertainty for the utility of the first decision is less than the lower bound of uncertainty for the utility of the second decision, then we may certainly remove the first decision from our collection of decisions that we need to consider in order to find the optimum.

Therefore, we proceed in stages. At each stage, we sample the function for a variety of values of $d$ that we have not yet rejected. We build an emulator based on this sample and we proceed to reject decisions using the above comparison, based on the current emulator. After a few stages, we typically have a good idea as to the maximum utility that we can obtain and the collection of decisions which we should consider that all achieve utilities close to this value.

We have described the problem as though there is a single optimum of interest. However, typically, we will want to investigate the tradeoffs between achievable values for each of the leading attributes. For example, we often want to maximise benefits and minimise costs, so that our concern may be to learn how much extra benefit may be achieved for different levels of cost.

Typically, therefore, the aim of a decision support analysis is to identify and invert the Pareto boundary for the problem, namely those decision utilities for which there is no other decision utility which is lower than or equal for all decision attributes of interest and strictly lower for at least one decision attribute. For example, if the attributes are cost and benefit, then a point lies on the Pareto boundary of the expected cost against expected benefit plot if there is at least one decision leading to this combination of expected cost and expected benefit, and there is no other decision leading to either a larger expected benefit for the same or smaller expected cost or a decision leading to a smaller expected cost for the same or larger expected benefit. We identify the Pareto boundary in stages in the same way that we identified our overall maximum utility in the earlier problem description, at each stage eliminating any decision which is dominated by at least one other decision according to the current stage emulator. Note that, as for simple utility optimisation, we both identify the Pareto boundary and invert the boundary (i.e. find the collection of decision choices that are on or near the boundary).

The Pareto boundary identifies what is achievable in the decision problem. When we have identified the part of the boundary which best reflects our objectives in the decision problem, we compare the collection of decisions which best achieve these objectives by study of their risk profiles and any additional features of the problem not encoded within the model.

While the above description is general, there will be particular features which may require care in the evaluation of the overall collection of decision choices. In particular, many decisions are of a sequential nature. We make an initial decision, observe the consequences and any other real world developments and then choose our second decision and so forth. In principle, the solution to such problems is well known and proceeds by backward dynamic programming. (A familiar example is the solution of a decision tree.) We cannot directly identify the best choice for the first decision, as its value depends on the choices for subsequent decisions, which we have not yet determined. Therefore, we solve the problem in reverse order. When we make the final decision, we know all of the previous decisions and observations, so we can choose the best final decision for each version of this information. As we now know the final decision, for each earlier choice, we can move back to the penultimate decision and identify the best choice for each version of the relevant information that we will have at this earlier time. Continuing in this way allows us to identify the best initial decision and to identify a strategy for choosing the best decision at each stage of the process, given earlier decisions and outcomes, though we will monitor this process over time and revisit later choices as they arise. For complex problems, determining such sequential strategies may be extremely computationally expensive. In such cases, there are good practical strategies to approximate the best overall solution based on building combined emulators for the multi-stage decision problem.

# 5. Tools for managing issues: governance and organisation

## Introduction

There are technical, methodological and statistical tools for managing issues arising from the use of modelling in complex organisations. There are also ways of addressing these issues through management, organisation and governance.

Ultimately, the appropriate use of modelling to support decisions is a question of sound methodology operating within a framework of good governance. We have addressed issues of sound methodology in the previous section. However, even the use of a sound methodology requires a process of assurance for decision makers. There is no point in using good techniques and methods if the decision makers cannot distinguish between sound and unsound analysis.

Of course, it is not just the decision makers in an organisation that require assurance about the use of models. External stakeholders are also important, particularly where the decisions are being made by a public authority, such as a regulator. If people's lives are being affected by the output of a model, they deserve some insight to how the modelling is being carried out and the underlying assumptions behind it.

We can consider the organisation and governance considerations as follows.

- Dynamics of modelling activity within an organisation.
- Models of governance – assurance to control.
- Quality assurance processes and internal audit. Building trust.
- The use of external modelling resources and capability
- Knowledge management, curation, change management and version control.
- Accountability and challenge in public decision making
- Communication challenges
    - Between expert analysts and decision makers
    - Between an organisation and external stakeholders

## Dynamics of the use of multiple models within organisation

As described previously, multiple models exist within a complex organisation for many reasons, both good and bad. The organisational challenge is that these models are both consistent and aligned with the organisation's objectives.

There are alternative ways of doing this.

> ***Through "business as usual" management structures, leadership and strategic management.*** If everyone understands the corporate objectives and how to achieve them and if management structures incentivise people to realise these objectives, then individual modelling activity should be aligned with these objectives. However, there is still a risk of inconsistency in approach and of input assumptions unless the culture of internal communication and collaboration is remarkably good.

***Supplement with corporate processes to promote quality assurance and consistency.*** *An internal QA process would ensure some integrity of modelling methodology and input data. Also, some sort of processes to agree a set of common assumptions for key variables, such as discount rate, energy prices, GDP growth and so on would help ensure that the modelling input to any decision or set of decisions were consistent. However, this approach should include a consideration of uncertainty, otherwise the model assumptions may be consistent, but would also be consistently wrong. Such processes might be supported by a central resource with specialist modelling expertise.*

***Centralise responsibility for modelling and analysis through a common resource.*** *This may make sense from an efficiency perspective, give modelling and analytical skills may be scarce and specialised, particularly in small and medium sized organisations. It would also help ensure consistency in approach and promote quality assurance. There would be a challenge in ensuring sufficient input from the specialist technical and market knowledge that may exists in the organisation and gaining sufficient "buy-in" from the sponsoring functions. A variant on this approach would be the "internal consultancy" model, where a central expert team is available to advise and support analytical teams in carrying out the work.*

***Outsource modelling capability to a third party (or parties).*** *Typically, the third party might be a commercial consultancy business. This may make sense from a cost and efficiency perspective. It is expensive both to maintain an internal modelling capability and to develop bespoke models. Public bodies may struggle to recruit and retain such staff due to differences in pay and benefits. For public bodies, in particular, the use of external modelling resource might also be viewed as conferring a greater degree of objectivity and independence. However, there are many challenges, not least with transparency and consistency when (as is typical) more than one external organisation is used, and with maintaining the necessary internal knowledge of the approaches used.*

It is over-simplistic to consider organisations as self-consistent entities united by a common purpose. In practice, individuals and teams within organisations will often have their own agendas. These will range from self-interest (personal advancement, gain or status) and team-based dynamics to well-intended, but differing views of organisational objectives and priorities, along with deeply embedded cultures of how modelling is used in certain applications or disciplines.

In public decision making, differing political/ economic perspectives and ideologies may also have an impact.

Management structures and processes are designed to overcome these different agendas and align the actions of individuals and teams to the overall goals and purpose of the organisation. However, this is not easy and is rarely, if ever, entirely successful.

Inevitably there will be tensions, particularly between decision makers, sponsoring departments and any centralised resource. It is important that these tensions are recognised and addressed. A central analytical capability or assurance function can also help, but can also become part of that dynamic, with its own agenda.

A surprisingly common approach is to allow different modelling insights from different interests to "compete" for the attention and influence of decision makers. This might be particularly true in public decision making, where external stakeholders might produce their own modelling evidence in response to a consultation process. But it can also be the case where there are competing interests within a complex corporate structure. Decision makers may feel, with some justification, there is merit in using diverse analysis from differing perspectives to inform their decisions. However, it would still be important for decision makers to be assured of quality and to understand differences in modelling approaches and input assumptions. Otherwise, there is a risk that the process simply becomes one of the decision makers choosing the model that provides answers that fit best with their own prior beliefs.

In short, there are no easy answers. But in even moderately complex organisations it is unlikely that leaving the problem of multiple models to "business and usual" management approaches, and decision maker "instinct" will inform high quality decision making. Decision making in such organisations would be likely to benefit from some sort of organisational solution or corporate process to promote consistency between models, transparency of approach and integrity of methodology.

## Knowledge management and models

Corporate knowledge is a valuable resource for any organisation, reflecting the totally of expertise, experiences, information and insight held by people and teams within an organisation. This knowledge can be paper based, held electronically or can exist solely in the knowledge, experience, networks and expertise of employees. Models are also repositories of knowledge, embedding an understanding of how the world works as well critical input data. The rationale for building a model in a particular way may not always be transparent but will reflect someone's considered (or less considered) choice based on their knowledge and expertise.

There is a general challenge in complex organisations of ensuring that corporate knowledge is captured, made available across the organisation and is maintained regardless of organisation change and turnover of staff. More typically, knowledge is lost when people move and teams are disbanded. Where knowledge and information is captured, it is often only available to a part of the organisation (silos), which may be unwilling to share ("knowledge is power"). Even where knowledge is made available more widely, it might not be in a form that allows it to be accessed easily.

All of this is true of modelling as much, if not more so, than other forms of data, knowledge and information. The corporate world is littered with models that nobody in the organisation understands or can use. Even worse, it is probable there are models being used to support decision making which are based on assumptions, understanding and approximations which have long been forgotten.

Good knowledge management has three components.

1. ***Culture***. Individuals and teams in an organisation need to understand the importance of good knowledge management and the value of building the corporate knowledge resource and sharing knowledge with others. This should be reinforced by leadership messages and the right behaviours should be rewarded and reinforced through performance incentives. The goal is for the organisation and individuals within it to consider knowledge as valuable resource to be built, nurtured, maintained and made useful.
2. ***Processes***. An organisation needs the right internal process and protocols to ensure that knowledge is captured and made available in a form that easy to access and use. Such processes would include protocols for storing data and papers in corporate systems. File naming protocols, the use of keywords, folder structures and so on need to be consistent across an organisation. They would also include processes for capturing knowledge from leavers and ensuring it is either transferred to others or stored in systems. These processes should be robust in the event of organisational restructuring.
3. ***Systems***. An organisation needs suitable information systems to allow authorised people to access knowledge, data and information easily and in useful form. Such systems should also be able to manage restricted information and ensure the information can only be changed or updated according to the corporate processes and protocols.

The same considerations apply to modelling as much as other forms of knowledge. Good management of models within an organisation requires the right cultural and managerial context, it requires consistent processes and protocols, particularly with respect to documentation and hand-over during organisational change, and it requires suitable information systems.

Good knowledge management can play a large part in engendering trust in models and supporting good decision making in an organisation. If the knowledge and assumptions embedded in models is widely accessible in an organisation, new knowledge can be incorporated, and assumptions can be challenged. People building models in related areas can ensure consistency, or at least be aware of where there may be inconsistencies. Input data can be shared between models and model outputs can be used with a fuller understanding of its limitations and degree of uncertainty.

## Communicating modelling results to decision makers

There is absolutely no point in undertaking modelling activities if the results cannot be effectively explained effectively to decision makers. Indeed, in such cases modelling might do more harm than good, by giving false assurance or leading decision makers to overlook other (non-quantified) evidence.

Effective communication of modelling evidence is not just a question of presenting the results. Decision makers also need to understand the uncertainty and robustness of the modelling evidence and its relevance to the decision. What are the results telling the decision maker, and how robust are those conclusions?

Culturally, this challenges the natural human desire for certainty and assurance, which means that analysts and decision makers are sometime inclined to understate the scale of uncertainty and focus on a single answer for a single variable (often expressed to an unfeasibly high degree of accuracy). The shortcomings of this approach should not need explaining.

The task of communicating the consequences of actions today in a way that recognises the uncertainty of outcomes, optionality and risk mitigation is not a trivial one. The results from modelling activities, particularly where uncertainty has been properly accounted for, can be statistically complex and far from easy for those unfamiliar with statistical concepts to understand. Also, in complex environments it is rare for there to be a single determinant of what outcome is the "best" outcome (such as NPV). At the very least there are likely to be trade-offs between expected value and risk/uncertainty. There are also likely to be trade-offs against other criteria (social equity, sustainability and so on), particularly in the public realm.

Furthermore, actions today may close-down or open-up options and choices at some later date, including the scope to manage risk by changing course should the outcomes not turn out to be as favourable as expected.

As with other areas, there are technical and [managerial] ways of bringing the full complexity of modelling outcomes to life for decision makers. Any research programme looking at the use of models in complex organisations will need to explore these challenges.

Options worth exploring include:

***Engagement with decision makers at every stage of the modelling process.*** *Decision maker should be engaged early so that modellers understand the basis on which a decision is likely to be made. Which variables are likely to be most important? What do the decision makers expect modelling to show them? This is important, both to ensure modelling adds value to the decision, but also to set expectations about what modelling can and cannot do.*

***Data visualisation techniques.*** *Developing narratives linked to key decision criteria to bring statistical data to life. Expressing model outcomes in plain language in a narrative that relates to the decision will make it easier for decision makers to understand exactly what a model is and is not telling them.*

***Scenarios.*** *Scenarios are sets of internally self-consistent background assumptions, which are representative of a specific narrative. It adds value to multi-variable sensitivities because it allows the decision makers to focus on a small number of representative background scenarios rather than an infinity of possible sets of input assumptions. It also defines these background scenarios in terms of a small number of key drivers (such as climate action, economic growth, political change and so on). As a narrative based approach, using scenarios supports the use of narratives to communicate statistically complex results.*

***Understanding how modelling evidence relates to other information available to decision makers.*** *Modelling evidence is rarely the most important factor in decision making, nor should it be. More typically, modelling either supports or challenges the qualitative judgement of decision makers, based on their experience and understanding of the world. Modelling and analysis can provide either assurance or allow refinement of decisions (e.g. exactly how much of x should we buy? or how big a y should we invest in?). It is important that modellers understand the decision framework and the role of modelling evidence within that.*

***Communicating risk, and its mitigation.*** *Demonstrating that outcomes can/ cannot be consistent with risk limits. Sometimes the most important role of modelling is to demonstrate that an organisation is functioning within financial risk limits or in not going to breach some other important criteria, such as environmental limits or safety. Finding ways of demonstrating the degree of assurance often an important part of communicating risk. Comparing the risk to other risks that decision makers might be more familiar with may be one way of giving statistical risk measures more salience (e.g. "the risk of x happening is less than the probability of you winning the lottery").*

***Communicating to external audiences.*** *Communicating outcomes of modelling, risk, and its relevance to decision making to external audiences is also important, particularly in the realm of public decision making. Many of the same considerations apply, but the challenge of addressing the gap between experts* *and those affected by a decision is even greater. This challenge has been illustrated starkly during the current COVID crisis, where the rationale for profound decisions supported by complex modelling evidence has had to be communicated to the wider public.*

# 6. Next steps and practical application

This paper has summarised issues inherent in the use of modelling within organisations from the perspectives of both technical modelling and its use in supporting decisions. There is much scope for further technical work by methodology specialists on uncertainty quantification for specific applications, and we would encourage such new activity in addition to the considerable advances already made in many fields.

We wish to highlight, however, the pressing need to develop routes for spreading good practice in use of modelling beyond specialist decision analysis methodology communities – there is very broad use of modelling across government and the private sector, and it is simply impractical for highly specialist external advisors to become involved in every single relevant analysis project.

There are already some efforts in this direction in government. Following publication of the HM Treasury Aqua Book [6], a cross-department team from central government produced a 'Uncertainty Toolkit for Analysts in Government' [23]. This provides guidance to the broad spectrum of analysts in the field on how uncertainty can be managed in relatively straightforward situations (for instance where uncertainties in model inputs are easily quantified, uncertainty is low dimensional so the uncertain space can be explored comprehensively, and there are no major issues of discrepancy between model structure assumptions and the real world system considered).

However, many real-world situations, particular looking at capital planning and long-term policy and strategy decisions, do not satisfy these conditions. It is thus necessary to provide practical guidance, both in general and for decision analysis questions in particular areas, on how organisations can perform useful analysis supporting good decision making for the actual real-world question under study. While there are many research papers and equivalents on these complex situations, we have not been able to identify much practical guidance for decision support teams in the field.

It is probably impractical to aim such resources at the full spectrum of analysts within organisations. The target audience might more naturally be more specialist analysts and teams, and those who take a strategic lead on analysis practice. The introduction of new practices will not happen overnight. It will be necessary to innovate for the medium to longer term while simultaneously providing support to decision makers in the short term.

All of this must be seen in the context that models do not take decisions, humans do. Models are just one part of decision analysis and decision support. One must remember that it is decision owners and not analysts who define the questions to be posed of modelling (a good summary of this is the question 'who decides what matters?'). Effective two way communication is needed between decision support teams and decision makers, so that the former have a proper understanding of the context of the decision question, and the latter have a proper understanding of what modelling analysis is and is not telling them.

This is an ambitious manifesto, and a challenging one to implement. We emphasise that the potential for economic and societal gains from improved decision analysis practice are very great, not least as the costs of improved analysis are typically small compared to the scale of the consequences of decisions.

Optimal decisions in model-world are of at most tangential relevance without assessing the relationship between the model and the real world. The ultimate aim is, therefore, to provide as good an understanding as possible of the consequences of decisions in the actual real world system in question based on what we know and can infer.

# Appendix 1: Review of the use of models in government

From time to time, Government issues guidance and reviews in relation to the use of modelling to support decision making and policy development. In contrast to the commercial sector, Government is accountable to wider society. It needs to provide assurance that modelling is being used appropriately to support decisions, given the scope for an adverse impact on people's lives.

In this review we have looked at the following reports.

2012 Blackett Review: high impact low probability risks.

2013 The Green Book: appraisal and evaluation in central government

2013 The Orange Book: Management of risk: Principals and Concepts

2015 Macpherson Report: Review of quality assurance of Government analytical models: final report.

2015 Macpherson: progress report

2015 The Aqua Book: Guidance on quality analysis for government

2015 Aqua Book resources

2018 Blackett Review: computational modelling future technologies

These reports cover a period in which the priorities of government, and governments themselves, have changed and methodologies have expanded. A focus on quality improvement and risk management has expanded into a more general attention to the benefits of modelling and, more recently, statistical ideas have also been enlarged into the area of uncertainty, or uncertainty quantification (UQ) and uncertainty management (UM). At the same time government is looking to "big data" for methodological advances, driven by some successes of machine learning and AI.

We simply list some of the point of agreement from the documents listed in Appendix 1. Some of the points will be revisited later sections.

*Decision support.* One concept runs clearly through all the documents, which is that the overriding need for modelling is to make good or better decisions. It is also clear that there are special but sometimes subliminal drivers, notable (i) to cut costs or, more positively, to obtain value for money in public investment (ii) to avoid or mitigate risks and hazards, both economic and physical.

*The purpose of modelling.* There is general agreement about these: (i) forecasting, (ii) monitoring, (iii) control/management, (iv) optimisation, (v) investment decisions. The first four can be considered shorter term.

*Fit for purpose*. There is general agreement that the best models, and the data collection and calibration that supports them, are designed for purpose. This throws up very topical issue of whether the drive to acquire and make use of the huge volumes of data in some fields can be tailored effectively for specific uses. Can patient's data be used to improve treatment? Can social media data be used to track epidemics? Can crime data be used to model the causes of knife crime?

*Different "sectors" of modelling*. It is recognised that historically the most complex models are physically based which are divided broadly in accordance with to the traditional scientific and engineering areas, but with good recognition of hybrid areas: biomathematics, multi-physics, environmental and so on. The gap between physical and economic modelling is highlighted, but not really elaborated. This also applies to modelling in other areas such social science and social policy. But it is recognised that modern decision support may need to cover several sectors. This is one type of "multiple model". The different sectors will often be driven by different traditions: physical laws described via differential equations in engineering; game theory, choice theory and theories of equilibrium in economics; social mobility and inequality and much else in the social sciences, huge advances in medical informatics, biomathematics and epidemiology.

*Computation.* From within compute science new, more algorithmic, styles of modelling and different data structures and types of models are fast developing: machine learning, images processing, multilevel (deep) models, Natural Language Processing. There is no clear message about what kind of future this will lead to, in terms of modelling.

*Uncertainty* . Whether couched in terms of the need for confidence intervals, simple sensitivity or a desire for a more holistic scenario-based approaches there is no doubt that the need to have some kind of uncertainty wrapping for modelling is now felt to be urgent. The uncertainty itself need to be assessed and models, presented in an easily understood way and with good graphics.

*Multiple models.* A system approach to modelling for decision making in the complex world of government must remain important. So, then, the system be should be reflected in the modelling environment. Different sectors of modelling, mentioned above, must be combined with different mathematical traditions and with great attention to uncertainty. Models can literally be joined, for example engineering and cost models in infrastructure investment: CAPEX and OPEX.